\documentclass[acus]{JAC2001}
\usepackage{graphicx}
\begin{document}

\thispagestyle{empty}

\onecolumn

\begin{flushright}
{\large
SLAC--PUB--9255\\
June 2002\\}
\end{flushright}

\vspace{.8cm}

\begin{center}

{\LARGE\bf
Tuning Knobs for the NLC Final Focus\footnote
{\normalsize{Work supported by
Department of Energy contract  DE--AC03--76SF00515.}}}

\vspace{1cm}

\large{
Y.~Nosochkov, P.~Raimondi, T.O.~Raubenheimer, A.~Seryi,
M.~Woodley \\
Stanford Linear Accelerator Center, Stanford University,
Stanford, CA 94309}

\end{center}

\vfill

\begin{center}
{\LARGE\bf
Abstract }
\end{center}

\begin{quote}
\large{
Compensation of optics errors at the Interaction Point (IP) is essential
for maintaining maximum luminosity at the NLC.  Several correction systems
(knobs) using the Final Focus sextupoles have been designed to provide
orthogonal compensation of linear and the second order optics aberrations
at IP.  Tuning effects of these knobs on the 250~GeV beam were verified
using tracking simulations.
}
\end{quote}

\vfill

\begin{center}
\large{
{\it Presented at the 8th European Particle Accelerator Conference 
(EPAC 2002)\\
Paris, France, June 3--7, 2002} \\
}
\end{center}

\newpage

\pagenumbering{arabic}
\pagestyle{plain}

\twocolumn

\title{
TUNING KNOBS FOR THE NLC FINAL FOCUS~\thanks
{Work supported by Department of Energy contract 
DE--AC03--76SF00515.}\vspace{-4mm}}

\author{
Y.~Nosochkov, P.~Raimondi, T.O.~Raubenheimer, A.~Seryi,
M.~Woodley \\
SLAC, Stanford, CA 94309, USA\vspace{0mm}}

\maketitle

\begin{abstract}

Compensation of optics errors at the Interaction Point (IP) is essential
for maintaining maximum luminosity at the NLC.  Several correction systems
(knobs) using the Final Focus sextupoles have been designed to provide
orthogonal compensation of linear and the second order optics aberrations
at IP.  Tuning effects of these knobs on the 250~GeV beam were verified
using tracking simulations.

\end{abstract}

\vspace{-1mm}
\section{INTRODUCTION}

\begin{figure}[htb]
\centering
\vspace{-3mm}
\includegraphics*[width=80mm]{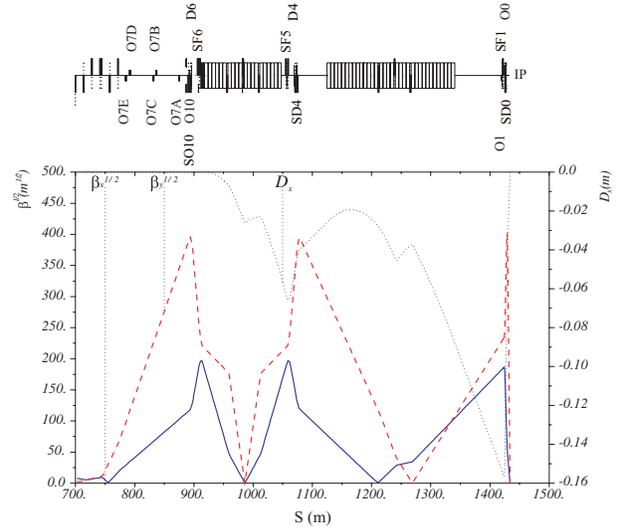}
\vspace{-3mm}
\caption{Optics and correctors in the NLC Final Focus.}
\label{fig:optics}
\vspace{-3mm}
\end{figure}

The NLC Final Focus (FFS) optics~\cite{ffs} is designed to produce a very
small vertical beam size of 3~nm at the Interaction Point (IP) in 500 GeV
{\it cms} collisions.  To attain these small spots, the NLC FFS design
includes a number of non-linear elements; the optics for the NLC FFS is
shown in Fig.~\ref{fig:optics}, where sextupoles are denoted with first
letter ``S'', octupoles with ``O'', and decapoles with ``D''.  These
correctors are used to compensate the high-order geometric and chromatic
effects generated in the nominal FFS optics as well as fold the beam halo
in on itself to reduce the number of large amplitude halo
particles~\cite{tuning}.

Unfortunately, alignment and field errors in the FFS are amplified by the
strong focusing and may lead to enlargement of IP beam size and subsequent
luminosity loss.  To maintain the maximum luminosity, aberrations
increasing the beam size at IP have to be corrected.  The correction
procedure envisioned for the NLC FFS is very similar to that implemented on
the Stanford Linear Collider (SLC) FFS or the Final Focus Test Beam (FFTB)
\cite{FFTB}:
\begin{enumerate}
\vspace{-1mm}
\item the quadrupoles are aligned using beam-based alignment techniques such as the shunting method,
\vspace{-1.5mm}
\item the sextupoles are aligned in a similar manner,
\vspace{-1.5mm}
\item trajectories are fit to verify the first-order optics and fix the phase advance between sextupoles,  
\vspace{-1.5mm}
\item the sextupoles are set to minimize the chromaticity,
\vspace{-1.5mm}
\item global tuning correctors (knobs) are used to tune both the first-order and the nonlinear corrections using luminosity measurements. 
\vspace{-1mm}
\end{enumerate}

In the following, we describe a set of simple correction system (knobs)
using a minimum number of magnets for independent control of individual IP
aberrations.  The knob correctors are located at optimum optical positions
close to the IP to minimize the extent of optics perturbation.  Several
knobs based on the FFS sextupoles have beed designed to correct linear and
the second order optical aberrations at the IP.

\vspace{-1mm}
\section{LINEAR KNOBS}
\vspace{-1mm}

Five knobs have been designed to correct the following linear aberrations
at the IP:  longitudinal position of horizontal ($x$) and vertical ($y$) beta
waist, $x$ and $y$ dispersion, and $x$-$y$ coupling.  
Correction of the first order aberrations requires adjustment of the normal
and skew quadrupole field in the FFS.  Similar to tuning techniques at the
SLC~\cite{slc} and the FFTB, we used $x$, $y$ offsets of the FFS
sextupoles to make these
knobs.  One can verify that a sextupole displaced by $\Delta x$ and $\Delta
y$ generates the following low order field on the reference orbit:
\vspace{-3mm}
\begin{eqnarray}
\Delta B_y=(-B^{\prime\prime}\Delta x)x + (B^{\prime\prime}\Delta y)y +
B^{\prime\prime}\frac{\Delta x^2-\Delta y^2}{2}, 
\label{eq:dBy} \\ 
\Delta B_x=(-B^{\prime\prime}\Delta x)y - (B^{\prime\prime}\Delta y)x +
B^{\prime\prime}\Delta x\Delta y, \;\;\;\;\;\;\;\;\;
\label{eq:dBx}
\vspace{-3mm}
\end{eqnarray}
where $B^{\prime\prime}\!=\!\frac{\partial^{2}B_y}{\partial x^2}$ is the
sextupole gradient.  Therefore, the feed-down normal and skew quadrupole 
strengths are 
\vspace{-1mm}
\begin{equation}
\Delta K_{1n}=-K_2\Delta x, \;\;\;\;\;
\Delta K_{1s}=K_2\Delta y, 
\label{eq:dK}
\vspace{-1mm}
\end{equation}
respectively, where $K_m\!=\!\frac{1}{B\rho}\frac{\partial^{m}B_y}{\partial
x^m}$ and $B\rho$ is the magnetic rigidity.  The last terms in
Eqn.~\ref{eq:dBy},~\ref{eq:dBx} lead to additional orbit, quadratic in
$\Delta x$, $\Delta y$, which may be not negligible for large offsets, however, it
is assumed that this orbit will be cancelled by the IP steering correctors.
Correction efficiency of the FFS sextupoles is enhanced by the large beta
functions and phase advance of $\Delta\mu^s\!=\!\pi/2\!+\!n\pi$ from IP.

\vspace{-1mm}
\subsection{Waist and Horizontal Dispersion}
\vspace{-1mm}

Enlargement of IP beam size and subsequent luminosity loss may be caused by
longitudinal displacement $\Delta s$ of focusing waist and residual horizontal
dispersion $\Delta \eta_x^*$ at IP.  The waist displacement increases IP
beta function as $\beta\!=\!\beta^*\!+\!\frac{\Delta s^2}{\beta^*}$, where
$\beta^*$ is the ideal beta at IP.  For maximum luminosity, waist position
has to be kept within $\Delta s\!\ll\!\beta^*$.

The beta waist and horizontal dispersion knobs have been constructed using
horizontal sextupole offsets.  According to Eqn.~\ref{eq:dK}, $\Delta x$
creates a normal quadrupole field which distorts $\beta_{x,y}$ and $\eta_x$
functions.  Beta perturbation propagates to the IP and for
$\Delta\mu^s\!=\!\pi/2\!+\!n\pi$ results in a longitudinal shift of the
beta waist.  The shift caused by a single sextupole is approximately
\vspace{-2mm}
\begin{equation}
\Delta s_{x,y} \approx \pm K_2L \Delta x \, \beta_{x,y}^s\beta_{x,y}^*
\cos{2\Delta\mu_{x,y}^s}\,,
\label{eq:ds}
\vspace{-2mm}
\end{equation}
where $L$ and $\beta^s$ are the sextupole length and beta function.  As
shown in Fig.~\ref{fig:optics}, the dispersion $\eta_x^s$ is not zero in most
of the FFS sextupoles, therefore $\Delta x$ also generates horizontal
dispersion at IP:
\vspace{-2mm}
\begin{equation}
\Delta \eta_x^* = K_2L \Delta x \, \eta_x^s
\sqrt{\beta_x^s\beta_x^*}\sin{\Delta\mu_x^s}.
\label{eq:dx}
\vspace{-2mm}
\end{equation}

Three sextupoles are required to construct three orthogonal knobs for the
$\Delta s_{x,y}$ and $\Delta \eta_x^*$ aberrations.  In each knob, the
sextupole offsets are varied linearly with the fixed scale factors to
produce a desired amplitude of the aberration.  The scale factors for
$\Delta x$ offsets are listed in Table~1.  Note that two sextupoles are
sufficient for $\Delta s_y$ and $\Delta \eta_x^*$ knobs for the following
reasons.  In the first case, the ratio $\eta_x^s/\sqrt{\beta_x^s}$ at SD0 and
SF1 is about constant which results in correction of $\Delta \eta_x^*$ when
$\Delta s_x$ is corrected.  In the second case, the $-I$ transformation
between SF5 and SF6 makes it possible to cancel $\Delta s_x$ and $\Delta
s_y$ with one scale factor.

\begin{table}[htb]
\vspace{-4mm}
\begin{center}
\caption{$\Delta x$ scale factors for $\Delta s_{x,y}$ and 
$\Delta \eta_x^*$ knobs.}
\medskip
\begin{tabular}{|l|c|c|c|c|}
\hline
\textbf{Sextupole} & \textbf{SD0} & \textbf{SF1} & \textbf{SF5} & 
\textbf{SF6} \\
\hline
$\Delta s_x$      & 1 & 0.6151 & 0 & -6.1249 \\
$\Delta s_y$      & 1 & 0.6151 & 0 &  0      \\
$\Delta \eta_x^*$ & 0 & 0      & 1 &  0.2609 \\
\hline
\end{tabular}
\end{center}
\vspace{-4mm}
\end{table}

MAD~\cite{mad} calculations show the following effects from these
knobs:  
$\Delta s_x\!\approx\!-3122\,\Delta x_{{\scriptscriptstyle SD0}}$, 
$\Delta s_y\!\approx\!-60.36\,\Delta x_{{\scriptscriptstyle SD0}}$, 
$\Delta \eta_x^*\!\approx\!-0.4981\,\Delta x_{{\scriptscriptstyle SF5}}$.  
It is expected that $x$, $y$ resolution of sextupole movers will be
close to 50~nm which is sufficient for accurate tuning of
$\Delta s_{x,y}$ and $\Delta \eta_x^*$ -- if desired, the magnet could be split onto
two movers which would effectively reduce the required resolution.

\vspace{-1mm}
\subsection{Vertical Dispersion and Coupling}

Orthogonal knobs to correct the vertical dispersion and betatron coupling at IP
have been constructed using vertical sextupole offsets.  The $\Delta y$
offsets create a skew quadrupole field which couples the $x$ and $y$
motion.  Vertical dispersion at IP, caused by a single sextupole, is given by
\vspace{-2mm}
\begin{equation}
\Delta \eta_y^* = -K_2L \Delta y \, \eta_x^s
\sqrt{\beta_y^s\beta_y^*}\sin{\Delta\mu_y^s}.
\label{eq:dy}
\vspace{-2mm}
\end{equation}

In general, betatron coupling is described by four orthogonal matrix terms
which can be independently tuned at IP using four skew quadrupoles located
at the following phase advance from IP:  $[\Delta \mu_x,\Delta \mu_y]\!=\!
[0,0],[\frac{\pi}{2},0],[0,\frac{\pi}{2}], [\frac{\pi}{2},\frac{\pi}{2}]$.
However, by design, all FFS sextupoles are located at
$\Delta\mu_{x,y}^s\!=\!\pi/2\!+\!n\pi$ from IP, and hence only one coupling
term can be created using the sextupoles.  Regardless, this is the dominant
coupling term in the FFS since most aberrations will be created by
quadrupoles with the largest beta functions which are at the same phase as
the sextupoles.  The remaining three coupling terms can be compensated
using three additional skew quadrupoles in the FFS system.  So far, it has
been found that the effect of the three minor terms is rather small and
there is a full 4-D coupling correction section upstream of the beam
delivery system.

Two sextupoles were used in the knobs for vertical dispersion and
coupling as shown in Table~2, where the scale factors for
$\Delta y$ are listed.  The SD0, SD4 offsets cancel vertical dispersion and
create coupling, while SF5 and SF6, located $-I$ apart, do the opposite.
The effect of $\Delta \eta_y^*$ knob is estimated to be
about $\Delta \eta_y^*\!\approx\!0.06918\,\Delta y_{{\scriptscriptstyle
SF5}}$.  This produces noticeable enlargement of vertical beam size 
when $\Delta y_{{\scriptscriptstyle SF5}}$ is a few microns.  The beam
sensitivity is much greater to the coupling knob, where an offset of
$\Delta y_{{\scriptscriptstyle SD0}}\!=\!0.5\,\mu$m increases the IP vertical
beam size by a factor of three.  This enlargement is roughly linear with
$\Delta y$ and it implies tight tolerances on the sextupole vertical
alignment.

\begin{table}[htb]
\vspace{-4mm}
\begin{center}
\caption{$\Delta y$ scale factors for $\Delta \eta_y^*$ and 
coupling knobs.}
\vspace{2mm}
\begin{tabular}{|l|c|c|c|c|}
\hline
\textbf{Sextupole} & \textbf{SD0} & \textbf{SD4} & \textbf{SF5} & 
\textbf{SF6} \\
\hline
$\Delta \eta_y^*$ & 0 & 0      & 1 &  0.2610 \\
Coupling          & 1 & 2.6490 & 0 &  0      \\
\hline
\end{tabular}
\end{center}
\vspace{-8mm}
\end{table}

\vspace{-1mm}
\subsection{Simulations}
\vspace{-1mm}

Spot size tuning using the linear knobs has been tested in tracking
simulations using DIMAD~\cite{dimad} and MATLAB-LIAR~\cite{matliar} codes.
The FFS design for 250~GeV beam was used where the unperturbed beam size at
IP is $\sigma_x^*/\sigma_y^*\!=\!243/3.00$~nm.  In the tests, the nominal
IP beam emittances were used for injection into the FFS rather than the
smaller design emittances which include allowances for tuning errors,
aberrations, and emittance dilutions.  Select errors were applied to the
FFS magnets which caused enlargement of the IP beam size, and then the
knobs were used to minimize this increase.  In a typical simulation, 2000
particles were tracked through the collimation and FFS sections with the
initial gaussian distribution in phase space and a double-horned
distribution for the energy spread.  Emittance growth due to synchrotron
radiation in the bending magnets and quadrupoles was included in the DIMAD
simulations which also causes a slight increase in the nominal IP beam
sizes.  The knob tuning was done manually in DIMAD and an automatic routine
was used in MATLAB-LIAR.  Two of the tests are presented below, although
more cases have been studied.

In the first test, a gradient error of $\frac{\Delta K_1}{K_1}\!=\!10^{-4}$
was applied to the final doublet quadrupole QF1.  According to MAD, this
results in IP waist offsets and horizontal dispersion as follows:
$\Delta s_x\!=\!-4.769$~mm, $\Delta s_y\!=\!0.134$~mm, and 
$\Delta \eta_x^*\!=\!44.8\,\mu$m.  
In this case, the $\Delta s_{x,y}$ values are comparable with the IP beta
functions $\beta_x^*/\beta_y^*\!=\!8/0.11$~mm.  Analytically, this
increases $\sigma_x^*$, $\sigma_y^*$ by 16\% and 58\%, respectively.  In
addition, $\Delta \eta_x^*$ enlarges the horizontal size by about
12\%.  To compensate this error, beta waist and horizontal dispersion knobs
were used in the simulations.  The results of this correction using DIMAD
are presented in Table~3, where $x^*$, $y^*$ are beam
offsets at IP.  The correcting sextupole offsets were: 
$\Delta x_{{\scriptscriptstyle SD0}}\!=\!0.48$, 
$\Delta x_{{\scriptscriptstyle SF1}}\!=\!0.295$,
$\Delta x_{{\scriptscriptstyle SF5}}\!=\!84.05$,
$\Delta x_{{\scriptscriptstyle SF6}}\!=\!32.34\,\mu$m.
The SF5, SF6 offsets are large enough to create a non-negligible orbit $x^*$
which could be corrected using the FFS steering correctors.
Similar results were obtained using MATLAB-LIAR. 

\begin{table}[htb]
\vspace{-5mm}
\begin{center}
\caption{DIMAD results (in nm) for 
$\frac{\Delta K_1}{K_1}\!=\!10^{-4}$ in QF1.}
\medskip
\begin{tabular}{|l|c|c|c|}
\hline
& $\mathbf{\frac{\Delta K_1}{K_1}\!=\!0}$ & \textbf{No corr.} &
  \textbf{Corr.} \\
\hline
$\sigma_x^*/\sigma_y^*$ & 253.4 / 3.18 & 319.1 / 4.65 & 254.8 / 3.19 \\
$x^*/y^*$               & -27.6/-0.002 & -33.0/-0.132 & -87.9/-0.005 \\
\hline
\end{tabular}
\end{center}
\vspace{-5mm}
\end{table}

\begin{figure}[htb]
\centering
\vspace{-4mm}
\includegraphics*[width=80mm]{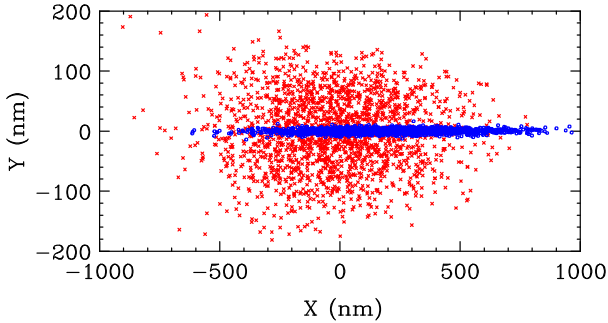}
\vspace{-2mm}
\caption{Particle distribution at IP before (red) and after (blue)
correction of $\Delta \theta\!=\!10^{-4}$ in QD0.}
\label{fig:qd0-rot}
\vspace{-3mm}
\end{figure}

In the second test, an $x$-$y$ rotation error $\Delta \theta\!=\!10^{-4}$
was applied to the final doublet quadrupole QD0. This generates two
dominant aberrations at IP: vertical dispersion and coupling. The results
of DIMAD simulations are listed in Table~4, and the correcting
sextupole offsets are:
$\Delta x_{{\scriptscriptstyle SD0}}\!=\!-2.87$, 
$\Delta x_{{\scriptscriptstyle SD4}}\!=\!-7.60$,
$\Delta x_{{\scriptscriptstyle SF5}}\!=\!-215$,
$\Delta x_{{\scriptscriptstyle SF6}}\!=\!-56.1\,\mu$m.
The corresponding IP distribution is shown in Fig.~\ref{fig:qd0-rot}.
Note that the uncorrected $\sigma_y^*$ is very large in this case which
implies tight tolerances on rotation errors in the final doublet and 
necessity for their compensation. 

More tests have been done using random rotation errors in all quadrupoles
in the collimation and FFS sections.  It has been found that the knob
correction is capable of reducing $\sigma_y^*$ enlargement to about 2\%
from an uncorrected size of 10$\sigma_y^*$.  For larger errors, the 
compensation is poorer, possibly due to effects such as chromatic
coupling and emittance dilution.

\begin{table}[htb]
\vspace{-5mm}
\begin{center}
\caption{DIMAD results (in nm) for 
$\Delta \theta\!=\!10^{-4}$ in QD0.}
\medskip
\begin{tabular}{|l|c|c|c|}
\hline
& $\mathbf{\Delta \theta\!=\!0}$ & \textbf{No corr.} & \textbf{Corr.} \\
\hline
$\sigma_x^*/\sigma_y^*$ & 253.4 / 3.18 & 258.6 / 64.0 & 253.2 / 3.40 \\
$x^*/y^*$               & -27.6/-0.002 & -26.1/-1.895 & 193.1/-0.0002 \\
\hline
\end{tabular}
\end{center}
\vspace{-5mm}
\end{table}

\vspace{-3mm}
\section{SECOND ORDER KNOBS}
\vspace{-1mm}

Taking into account the small $\sigma_y^*$ at the NLC, enlargement of IP
beam size due to high-order aberrations may also be significant.  The FFS
sextupoles, octupoles and decapoles may be used to create non-linear
correction knobs.  Since all of the FFS sextupoles are located at the same
phase from IP ($\pi/2\!+\!n\pi$), they create only limited number of second
order aberrations at IP.  At present, four approximately orthogonal knobs
have been constructed using six variable sextupole strengths to generate
the projected second order terms at IP:  $T_{122}$, $T_{166}$, $T_{342}$
and $T_{364}$.  Note that $T_{122}$, $T_{166}$ affect $\sigma_x^*$ and
$T_{342}$, $T_{364}$ change $\sigma_y^*$.  The knobs have been designed
using MAD matching routine.  Their effect can be analytically estimated by
constructing the thin lens second-order matrix containing the desired terms
and applying it to the IP phase space.

\begin{figure}[htb]
\centering
\vspace{-4mm}
\includegraphics*[width=80mm]{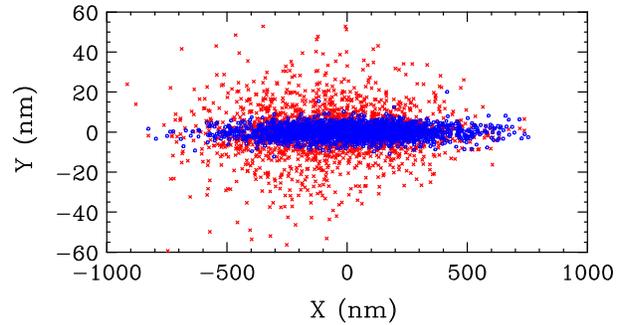}
\vspace{-2mm}
\caption{Particle distribution at IP before (red) and after (blue)
correction of $\frac{\Delta K_2}{K_2}\!=\!10^{-2}$ in SD0.}
\label{fig:sd0-dk}
\vspace{-3mm}
\end{figure}

An example of the second-order correction is presented in Table~5 and
Fig.~\ref{fig:sd0-dk}, where the strength error $\frac{\Delta
K_2}{K_2}\!=\!10^{-2}$ was applied to the SD0 sextupole.  MATLAB-LIAR
simulations also showed good compensation in another test, where random
strength errors were applied to five FFS sextupoles with {\it rms} value of
$\frac{\Delta K_2}{K_2}\!=\!3\!\cdot\!10^{-3}$.

\begin{table}[htb]
\vspace{-5mm}
\begin{center}
\caption{DIMAD results (in nm) for
$\frac{\Delta K_2}{K_2}\!=\!10^{-2}$ in SD0.}
\medskip
\begin{tabular}{|l|c|c|c|}
\hline
& $\mathbf{\frac{\Delta K_2}{K_2}\!=\!0}$ & \textbf{No corr.} & 
  \textbf{Corr.} \\
\hline
$\sigma_x^*/\sigma_y^*$ & 253.4 / 3.18 & 260.1 / 6.02 & 254.0 / 3.22 \\
$x^*/y^*$               & -27.6/-0.002 & -50.5/-0.035 & 9.6/-0.0003 \\
\hline
\end{tabular}
\end{center}
\vspace{-5mm}
\end{table}

\vspace{-3mm}
\section{CONCLUSION}
\vspace{-1mm}

A set of knobs based on the FFS sextupoles have been designed to correct
linear and second-order aberrations at the IP.  These knobs have been tested in
numerical simulations for several FFS errors and satisfactory
compensation has been achieved.  Additional high-order knobs are under study.
Further simulations will be needed to investigate the error range for an
acceptable knob correction and to develop a practical strategy for the beam
measurement and knob compensation.


\vspace{-1mm}
\begin{thebibliography}{9}
\vspace{-1mm}
\bibitem{ffs} P.~Raimondi, A.~Seryi, Phys. Rev. Let., 86, p. 3779, 2001.
\vspace{-1mm}
\bibitem{tuning} P.~Raimondi {\it et al.}, SLAC-PUB-8895, 2001.
\vspace{-1mm}
\bibitem{FFTB} P.~Tenenbaum {\it et al.}, SLAC-PUB-6770, 1995.
\vspace{-1mm}
\bibitem{slc} P.~Raimondi {\it et al.}, SLAC-PUB-7955, 1998.
\vspace{-1mm}
\bibitem{mad} MAD, http://wwwslap.cern.ch/mad/.
\vspace{-1mm}
\bibitem{dimad} NLC version of DIMAD,
http://www-project.slac.stanford.\\
edu/lc/local/AccelPhysics/Accel\_Physics\_index.htm.
\vspace{-1mm}
\bibitem{matliar} MATLAB-LIAR,
http://www-project.slac.stanford.edu/lc/\\
local/AccelPhysics/Accel\_Physics\_index.htm.

\end{thebibliography}
\end{document}